# Phase Behavior under a Non-Centrosymmetric Interaction: Shifted Charge Colloids Investigated by Monte Carlo Simulation

Luis E. Sánchez-Díaz[†], Chwen-Yang Shew[‡,*], Xin Li[†], Bin Wu[†], Gregory S. Smith[†] and Wei-Ren Chen[†,*]

[†]Biology and Soft Matter Division, Oak Ridge National Laboratory, Oak Ridge, Tennessee 37831, USA.

[‡]Department of Chemistry, College of Staten Island, City University of New York, Staten Island, New York 10314, USA.

Corresponding author's email address: chenw@ornl.gov, chwenyang.shew@csi.cuny.edu

ABSTRACT: Using Monte Carlo simulations, we investigate the structural characteristics of an interacting hard sphere system with shifted charge to elucidate the effect of the non-centrosymmetric interaction on its phase behavior. Two different phase transitions are identified for this model system. Upon increasing the volume fraction, an abrupt liquid-to-crystal transition first occurs at a significantly lower volume fraction in comparison to that of the centro-charged system. This is due to the stronger effective inter-particle repulsion caused by the additional charge anisotropy. Moreover, within the crystal state at higher volume fraction, the system further undergoes a continuous disorder-to-order transition with respect to the charge orientation. Detailed analyses in this work disclose the nature of these transitions, and orientation fluctuation may cause non-centrosymmetric unit cells. The dependence of crystal formation and orientational ordering on temperature was also examined. These findings indicate that the non-centrosymmetric interaction in this work results in additional freedoms to fine-tune the phase diagram and increase the functionalities of materials. Moreover, these model studies are essential to advance our future understanding regarding the fundamental physiochemical properties of novel Janus colloidal particles and protein crystallization conditions.

## I. Introduction

One grand challenge and opportunity in material sciences is to develop new materials that operate at high level of performance. Often the fundamental breakthroughs require *de novo* materials design at the atomic/molecular scale. In this pursuit, so-called "soft" materials offer many advantages. In general, they are multicomponent systems characterized by extensive chemical diversity. Because of this complexity, the number of phases and equilibrium structures potentially possessing different functionality is dramatically enriched in comparison to that of pure materials.[1,2] In most of soft matters, the constituent particle undertakes the centrosymmetric structure, in particular, from the theoretical standpoint. This work intends to cross this general view into a different horizon where soft matters are composed of non-centrosymmetric particles.

As a matter of fact, non-centrosymmetric interactions are of the important fundamental physics, which have profoundly linked with numerous material applications. The non-centrosymmetric interaction in crystals is the key to drive their nonlinear optical properties.[3] In soft matters, such as proteins and colloids, the significant non-centrosymmetric interaction is manifested in the form of asymmetric charge distribution. The asymmetric charge distribution along the membrane protein molecule creates the cross-membrane electrical potential, relevant to the membrane transport properties of cellular materials and cell adhesion, etc.[4] For colloid solutions, asymmetric interactions may account for the mysterious like-charge attraction even in dilute solution.[5] These studies focused on investigating the effect of non-centrosymmetric interactions in a chosen physical state.

From the framework of the non-centrosymmetric interaction, a greater challenge in condense matters, beyond the above studies, is to understand how the transition occurs among the physical states of substances in the presence of such an interaction. To facilitate future molecular designs, it is instructive to develop a simple theoretical model that can reveal the role of the non-centrosymmetric interaction on the phase behavior of materials. Among available substances, colloids are an ideal candidate to pursue the physics behind this topic.

Colloids are one important member of the soft matter family. They are complex fluids consisting of globular particles suspended in a liquid solvent. Their phase behavior and equilibrium structure are in principle determined by the inter-particle interaction. Gaining control over colloidal bulk properties through the design of the uniform inter-particle interactions has been an active research field of colloidal science.[6] Recent advances in material synthesis have allowed the creation of biphasic colloidal particles, which possess two compositionally different types of groups located at the particle periphery.[7-9] One remarkable feature of this new generation of colloids is that the non-uniform surface renders an additional degree of freedom for creating novel colloid-based materials.[10,11] This structural heterogeneity breaks the colloidal centrosymmetry and therefore results in a directional inter-particle interaction that depends not only on the translational coordinates, but also on relative colloidal orientation.

One of these biphasic particles is the Janus dendrimer.[12-15] It is created by grafting two types of terminal groups onto the surface of one single dendrimer.[16-18] This new kind of dendrimer has attracted considerable attention because of its role in the variety of industrial applications.[13-15] When dissolved in solvent, the different surface interactions lead to an anisotropic inter-dendrimer interaction. For example, it is now possible to synthesize a dendrimer consisting of two hemispheres with amine and hydroxyl terminal groups respectively. In an acidified aqueous environment, the excess protons in the solvent can charge the amines present in one hemisphere while the hydroxyls present in the other remain neutral. The unbalanced charge distribution leads to an asymmetric inter-dendrimer repulsion. Besides flexible synthetic strategies to construct these colloids, there exist simple theoretical models, for instance DLVO theory, to elucidate the behavior of colloidal solutions. There has been much interest in understanding the phase behaviors of symmetric repulsive Yukawa particles, the electrostatic potential in the DLVO model.[19-20] By shifting the charge site away from the center of a colloid in the DLVO model, the colloid captures the basic feature of "broken symmetry" that represents the simplest non-centrosymmetric model for colloids. In addition to its simplicity, the so-chosen model severs as guidance for the future synthesis of Janus colloids as well as for predicting the corresponding phase behavior. Investigating the general structural properties of particles interacting via anisotropic interaction is the motivation of this study. In our study, we investigate the dependence of phase behaviors on charge asymmetry using Monte Carlo (MC) simulations via benchmarking the results of reference symmetric systems. We describe the details of simulation in the following section.

**II. Monte Carlo Simulations**

Our modelled system is schematically represented in Figure 1. The interaction between the simulated colloidal particles is described by a hard sphere repulsion and a repulsive Yukawa repulsion whose center of charge is systematically shifted from the geometric center (Figure 1(a)) toward the particle periphery along the radial direction (Figure 1(b)). The reference system of Yukawa colloid, which is characterized by a centrosymmetric charge distribution, is denoted as CS in this report. As demonstrated by Figure 1(b), the colloid with asymmetric charge distribution is characterized by a nontrivial separation ($r_s$) of the charge center (CC) from geometrical center (GC). Mathematically, the pair-wise interaction potential can be cast as Eqn. (1).

$$u(r) = \begin{cases} \infty, & \text{if } r < \sigma \\ \Gamma \dfrac{\exp\left[-\kappa\sigma\left(\dfrac{r'}{\sigma}-1\right)\right]}{r'}, & \text{if } r \geq 0 \end{cases} \quad (1)$$

The infinite value of the potential energy when the distance $r$ between GCs of two particles is smaller than their diameters ($\sigma$) prevents the constituent particles from overlapping. At other particle separation, the interaction is a screened Coulomb repulsion, whose magnitude is collectively determined by screening length $\kappa$, interaction strength $\Gamma$, and reduced spacing between CC's, $r'$, that is normalized against $\sigma$.

Metropolis algorithm based Monte Carlo (MC) simulation is employed to obtain the equilibrium configuration of the aforementioned particles as a function of $r_s$, $\kappa$, $\Gamma$ and volume fraction $\phi$. In the course of simulation, the number of particle $N$, cubic simulation box size, and temperature are kept constant through applying NVT ensemble and periodic boundary condition.[21] The thermodynamic properties, such as internal energy and pressure, as well as pair distribution function $g(r)$ are carefully monitored to identify the equilibrium status of the simulated system. In most of the presented cases, the simulation is run on 10000 particles. Nevertheless, representative simulations are repeated to investigate the so-called size effect of the simulation box. Our results indicate that 10000 particles are indeed sufficient to deliver statistically reliable physics for the studied system.

**III. Results and Discussions**

The general description of their phase behavior usually begins with the consideration of phase transition. Indications of phase transition include the discontinuous transition of pressure and abrupt variation of heat capacity. In our simulation we monitor the evolutions of both physical quantities to elucidate the different phase behavior between centrosymmetric and asymmetric particles. First we calculate the excess heat capacity at constant volume $Cv$. This thermodynamic quantity is related to the fluctuation of potential energy, and is sensitive to the structure of condensed phases. As shown in Figure 1, we model the unbalanced intra-particle charge distribution by simply shifting it from the centre of mass towards the periphery. Figures 2(a) and 2(b) display the variation of the simulated reduced excess heat capacity per particle $Cv/Nk_B$ with a packing fraction $\phi$ for centrosymmetric particles and for the shifted charge particles. The degree of charge shifting is given by the ratio of shifted distance from the particle center $r_s$, to the particle diameter $\sigma$. We investigate two shifted charge systems with $r_s/\sigma = 0.1$ and 0.3. In order to benchmark the existing results of reference system ($r_s/\sigma = 0$),[19-20] in our Yukawa potential model the magnitude of repulsion $\Gamma$ is set to be 81 $k_B$T. The screening length $\kappa$ are set to be $6/\sigma$ and $10/\sigma$ respectively. For low packing fractions ($\phi < 0.06$ for $\kappa = 6/\sigma$ and $\phi < 0.09$ for $\kappa = 10/\sigma$) all three cases exhibit nearly the same $Cv$, indicating similar liquid structure among three different models. By increasing $\phi$, an increase of $Cv$ is observed because inter-particle interactions are progressively enhanced. As $\phi$ is increased further, the shifted charge model with $r_s/\sigma = 0.3$ shows a peak (for $\kappa = 6/\sigma$ at near $\phi = 0.07$ and for $\kappa = 10/\sigma$ at near $\phi = 0.093$) due to the transition from liquid to crystal state, and the crystal takes the form of FCC (face-center-cubic) structure based on the equilibrium trajectories. For the shifted charge model with $r_s/\sigma = 0.1$ such a transition is observed at near $\phi = 0.13$ for $\kappa = 6/\sigma$ and $\phi = 0.17$ for $\kappa = 10/\sigma$, and at $\phi = 0.21$ for the centrosymmetric model. The distinct $\phi$ at the transition point and the $Cv$ values after passing the transition point among these three cases may be attributed to their subtle crystal structures. We provide more details about the evolution of $Cv$ are given in Appendix A. In Figures 2(c) and 2(d) we present the evolution of pressure $P$ as a function of $\phi$ obtained from the equation of state calculations. It is clearly seen that the identified phase transition boundary is in a quantitative agreement with that determined from Figures 2(a) and 2(b).

Figure 3 gives the calculated phase diagram based on the variation of $Cv$ and P. Given the same repulsion, upon increasing the ratio of $r_s/\sigma$ from 0 to 0.3, the fluid-to-FCC crystal boundary is seen to progressively move towards lower $\phi$. One perspective to understand this observation is through the averaged pair potential energy $<U(r)>$ given in the inset. We briefly outlined the calculation of $<U(r)>$ in Appendix B. By increasing the parameter $r_s$ in our model, it is clearly seen that the average repulsive potential increase steadily. Crystal formation is therefore facilitated by the greater effective particle size due to a higher average repulsive potential. From the perspective of thermodynamics, this observation is resulting from the fact that the spatial randomness of the liquid phase driven by entropy is impeded by the higher energy due to intra-particle rotational degrees of freedom. It is important to point out that our model is conceptually different from the dipolar model[24] in which two oppositely charged sites (or domains) are placed onto the surface of a particle. The attractive interaction is the essential component in the dipolar model. An increase in the polarity, which can be achieved by increasing the separation between the positively and negatively charged sites in a particle, leads to a greater dipole moment. As a result, the effective attractive force between dipolar colloidal particles is enhanced and the repulsive force is attenuated.

In addition to monitoring the variations of $P$ and $Cv$, we have also employed other methods to phase transition. First in Figure 4 we present the MC-calculated inter-particle structure factor $S(Q)$ for particles with shifted charge of $r_s/\sigma = 0.3$ along with that of the reference centrosymmetric system of $r_s/\sigma = 0$. The $S(Q)$ for three states of a, b and c along the dotted line of Figure 3 are presented. In state a, the fact that both systems of $r_s/\sigma = 0$ and 0.3 are in the liquid state is clearly reflected by their corresponding $S(Q)$. In state b, the reference system of of $r_s/\sigma = 0$ is seen to remain in the liquid state but with higher local ordering judging from the height of the first interaction peak of its $S(Q)$. The formation of crystal for the system with shifted charge of $r_s/\sigma = 0.3$ is clearly revealed by the occurrence of the characteristic peaks in its corresponding $S(Q)$ (blue curve) presented in Figure 4. In state c, the $S(Q)$ presented in Figure 4 suggests that both systems of $r_s/\sigma = 0$ and 0.3 are in crystalline states.

In addition to the two-particle pair spatial correlation functions such as $S(Q)$, more structural information can be obtained by the $n$-particle spatial correlation function (with $n > 2$). For example, one can obtain the additional information about local oriental order from the three-point bond angle distribution function (BADF)[25-26]. First, we define bonds as the lines connecting nearest neighboring particles. A tagged particle $i$ together with two nearest neighbors $j$ and $k$ forms a bond angle with respect to $i$. The angle $\theta$ between the bonds joining the tagged particles with two neighboring particles is defined as

$$\theta = \cos^{-1}\left[\frac{\left(\vec{R}_{ij}\cdot\vec{R}_{jk}\right)}{|R_{ij}||R_{jk}|}\right] \quad (2)$$

In our calculation of this three-point spatial correlation function, we only consider the nearest neighbors lying within the first coordinate shell of the tagged particle. We calculate the distribution of the angle $\theta_c$ from 0 to $\pi$ and normalize the distribution in such a way that the probability of a bond angle between $\theta$ and $\theta + d\theta$ satisfies the following equation[26]

$$\frac{1}{\pi - \theta_c}\int_{\theta_c}^{\pi} P(\theta)d\theta = 1 \quad (3)$$

Figure 5 give the calculated $P(\theta)$ for states a, b and c along the dotted line gives in Figure 3. In their liquid state both $P(\theta)$ for $r_s/\sigma = 0$ and 0.3 are characterized by similar qualitative features. Because of the excluded volume effect, both $P(\theta)$ vanish when $\theta$ is less than 36°. Moreover, the peak at $\theta$ around 60° indicates that the equilateral triangle is indeed an energetic favorable configuration for the local packing of the triplet. However, the main peak of $r_s/\sigma = 0.3$ (blue curve) is seen to be diminishing in comparison to that of $r_s/\sigma = 0$ (red curve). This observation suggests that the local arrangement of equilateral triangular configuration is disturbed by the charge asymmetry. Upon increasing $\phi$ from state a to b, the sharpening maxima and minima clearly originates from the suppression of the local positional fluctuations. One noticeable feature for the case of $r_s/\sigma = 0.3$ (blue curve) is the appearance of additional peaks within the range of 90° < $\theta$ < 180°. Their presence signifies the formation of crystalline state.

As mentioned in the context of this report, we believe that the different evolution of $Cv$ values after passing the transition point among these three cases is caused by their subtle crystal structures. This argument is supported by the difference of $S(Q)$ and $P(\theta)$ shown in Figures 4 and 5. Their correlation is beyond the scope of this paper and work towards this end is currently ongoing.

Moreover, besides the liquid-to-crystal transition, we also notice that the shifted charge sites of asymmetric particles gain the orientation ordering as $\phi$ is further increased within the crystal phase. To study the orientational behavior of the asymmetric particles we calculate the orientational order parameter $<S>$ and correlation length $\xi$ as the index parameters.

We define the orientational order parameter $<S>$ as[22]

$$\langle S \rangle = \frac{1}{2} \langle 3\cos^2 \Delta\theta - 1 \rangle \tag{4}$$

where $\Delta\theta$ is the angle between the vectors of charge shift of two sampling particles, with respect to a pre-determined direction $z$, as shown in Figure 6. The vector $\vec{r}_{is}$ is defined by the position of the center mass and the distance and direction of charge shift $r_s$ for particle $i$. For a liquid in which its orientational correlation is completely random, the average of the cosine terms is zero and its $<S>$ is equal to zero. On the contrary, for a crystal with a perfect orientation, the corresponding $<S>$ is equal to one. Typical values for $<S>$ of a liquid crystal range between 0.3 and 0.9, depending on the ambient and the particle motion. In addition, we also calculate the correlation length of the orientational ordering $\xi$ based on the following equation

$$f(r) = e^{-r/\xi} \tag{5}$$

where $f(r)$ is the exponential function used to fit the peaks of correlation function $\langle \vec{r}_{is} \cdot \vec{r}_{js} \rangle$, which is calculated following:

$$\langle \vec{r}_{is} \cdot \vec{r}_{js} \rangle = \frac{1}{N(N-1)} \langle \sum_{\substack{i,j=1 \\ i \neq j}}^{N} (\vec{r}_{is} \cdot \vec{r}_{js}) \cdot \delta(\vec{r} - \vec{r}_i + \vec{r}_j) \rangle \tag{6}$$

One example of such a calculation for the particles with $r_s/\sigma = 0.3$ is given in Figure 7. From the definitions of $<S>$ and $\xi$ given above, they can be considered as index parameters for the occurrence of orientational ordering. In Figure 8 we give the calculated results of <S> and $\xi$ with $\phi$ (along the dashed line of Figure 3) for both cases of $r_s/\sigma = 0.1$ (blue symbols) and 0.3 (red symbols). Upon increasing $\phi$, it is clearly seen that the magnitudes of both <S> and $\xi$ increase progressively. This observation provides the evidence for the existence of the orientational ordering state within the FCC crystalline phase. The phase boundary is defined as the order parameter <S> is greater than 0 or the correlation length $\xi$ is greater than 1. Moreover, the volume fractions $\phi$ where the orientational ordering transition takes place revealed by the evolutions of <S> and $\xi$ are seen to be consistent with each other.

On top of the liquid-to-crystal phase diagram, the phase boundary of orientational transition is incorporated and the results are given in the top panel of Figure 4 for an asymmetric particle system ($r_s/\sigma = 0.3$). Figures 10 and 11 compare phase diagrams ($1/\kappa\sigma$ vs. $\phi$) and ($1/T^*$ vs. $\phi$), respectively, between both centroymmetric ($r_s/\sigma = 0$) and asymmetric particles ($r_s/\sigma = 0.1$ and 0.3). Similar to that of the liquid-to-crystal transition, the physical origin of this orientational ordering transition is the result of the compromise between energy and entropy. Upon increasing $\phi$, the parallel arrangement of the shifted charge vectors between the neighboring particles is seen to increase progressively as shown in Figure 7. Therefore, at a certain $\phi$, the energy decreases because this orientational ordering exceeds the loss of entropy and therefore results in the formation of orientational ordering state. The average size of a single domain and the degree of orientational ordering within which can be qualitatively estimated from the evolutions of $\xi$ and <S> given in Figure 8. As well as the difference between the orientational ordering boundary, the quantitative difference between the for the cases $r_s/\sigma = 0.1$ and 0.3 shown in Figures 2, 3, 10 and 11 clearly results from the two contending thermodynamic factors.

Various pair correlation functions displayed in the middle panel of Figure 9 provide insights into the structure of asymmetric particles in different physical states, and the corresponding 2D schematic pictures are shown in the bottom panel of Figure 9. Basically, both centrosymmetric and asymmetric particles pack based on their charge sites. In liquid state, $g_{cc}(r)$ (between the charge sites of asymmetric particles ($r_s/\sigma = 0.3$)) agrees well with $g(r)$ (between centrosymmetric particles ($r_s/\sigma = 0$)), due to the similar length scale of the effective electrostatic diameter for individual particles. The discrepancy between $g_{cc}(r)$ and $g_{gc}(r)$ (between geometric centers) of asymmetric particles is attributable to the orientation degree of freedom of the geometric center within a particle in the liquid and crystal state prior to the formation of the orientation ordering state. The general picture of phase transitions (Figure 11) can now be drawn as follows. During crystal formation, both types of particles manifest the abrupt transition because of the competition between electrostatic energy and entropy. In the crystal state, the non-symmetric hard core repulsion of asymmetric particles competes and particles start to align. The orientation ordering takes the continuous transition due to the athermal hardcore repulsion. Note that the local non-centrosymmetric crystal unit cell structure may be induced with orientation fluctuation of geometric centers before the highly order orientation state is reached at high packing fraction. Under such a condition, both charge sites and geometric centers tend to pack orderly, and the $g_{cc}(r)$ and $g_{gc}(r)$ of asymmetric particles coincide with that of the $g(r)$ of centrosymmetric particles.

**IV. Conclusions**

In conclusion, using MC simulation we investigate the phase behavior of asymmetrically charged particles. A liquid-to-FCC crystal and an orientational ordering transition within the FCC phase are identified. Furthermore, this work discloses the distinct nature between first-order-like liquid-crystal formation and continuous orientation ordering transition of this class of model colloids. In comparison to the centrosymmetric reference system, the charge shift introduces an additional degree of freedom, which alters the equilibrium condition of the system. We attribute the observed characteristic phase behavior to the thermodynamic competition between the potential energy and configurational entropy caused by the charge shift, a simple chemical modification.

Our finding provides a simple, but important, chemical architectural parameter in fine-tuning the boundary of crystal formation and orientation ordering of non-centrosymmetric charged colloidal particles. Additionally, our phase envelope is computed on the plane of $1/\kappa\sigma$ vs. $\phi$, which gives predictions of the solution condition, such as the range of the Debye-screening length, to achieve the liquid-crystal phase transformation for particles of different chemical architectures. (Note that the dependence of crystal formation and orientational ordering on temperature is shown). Moreover, it is worth mentioning a possible link between our computational results and experimental finings. It has been a common practice for facilitating crystallization of proteins from solutions via tuning the buffer condition and pH value.[27-30] In general, proteins are more complex than like-charged colloids dissolved in aqueous solutions.[31] Therefore, from the perspective of this work, it would be interesting to investigate how the possible anisotropic charge distribution on a protein surface, modulated by the variation of the saline condition, affects the protein crystallization from solutions. Besides, the orientational ordering of the shifted charge particles may afford a molecular pathway for a colloidal particle to interact with applied electric fields for which the electric polarization of crystal phase is induced. Such a possible material electric property will be further examined in a future simulation or by experimentally probing synthetic systems, such as Janus colloids.

**Acknowledgement**

This work was supported by the U.S. Department of Energy, Office of Basic Energy Sciences, Materials Sciences and Engineering Division. This Research at the SNS at Oak Ridge National Laboratory was sponsored by the Scientific User Facilities Division, Office of Basic Energy Sciences, U.S. Department of Energy.

**Appendix A. Evolution of *Cv* and Its Dependence on Charge Asymmetry**

The adopted principle of mapping out the phase diagram for the broken centro-symmetric particles is to identify the boundary where a phase transition occurs. This transition is reflected by the change in the configurational heat capacity at constant volume *Cv* as the volume fraction increases with fixed parameters $\kappa$ and $\Gamma$ in the potential model of Eqn. (1).

The configurational heat capacity for monoatomic fluids, which can be related through the fluctuation-dissipation theorem for the heat capacity at constant volume in the canonical ensemble, is given by the following expression

$$C_V^{total} = \frac{3}{2} N k_B T + \frac{\langle E^2 \rangle - \langle E \rangle^2}{k_B T} \tag{A1}$$

Where $N$ is the number of particles, $k_B$ is the Boltzmann's constant and $E$ is the internal energy, and the angle brackets represent the ensemble average. The configurational heat capacity at constant volume $Cv$ therefore is given by[21-23]

$$C_V = C_V^{total} - \frac{3}{2} N k_B T \tag{A2}$$

Our finding shows that $Cv$ tends to increase with $\phi$ for all cases except at the transition point; namely, . Using the Yukawa potential, this observation can be justified by the following. The variation of internal energy of the system $U$ can be expressed as

$$dU = \left(\frac{\partial U}{\partial T}\right)_V dT + \left(\frac{\partial U}{\partial V}\right)_T dV = C_V dT + \pi_T dV \tag{A3}$$

Where $T$ is the temperature, $V$ is the volume of the system, $C_V$ is the heat capacity at constant volume, and $\pi_T$ is the internal pressure. $\pi_T$ is known to be sensitive to the interaction energy. We can write $\pi_T$ as

$$\pi_T = \left(\frac{\partial U}{\partial V}\right)_T = \left(\frac{\partial U}{\partial \phi}\right)_T \left(\frac{\partial \phi}{\partial V}\right)_T \tag{A4}$$

Also, since $\left(\frac{\partial C_V}{\partial V}\right)_T = \left(\frac{\partial \pi_T}{\partial T}\right)_V$, it is found that

$$\left(\frac{\partial C_V}{\partial \phi}\right)_T \left(\frac{\partial \phi}{\partial V}\right)_T = \left(\frac{\partial \pi_T}{\partial \kappa}\right)_V \left(\frac{\partial \kappa}{\partial T}\right)_V \tag{A5}$$

Since $\phi = \pi \sigma^3 / 6V$, using Eqns. A4 and A5, we find

$$\left(\frac{\partial C_V}{\partial \phi}\right)_T \left(\frac{\partial (\pi\sigma^3/6V]}{\partial V}\right)_T = \left(\frac{\partial}{\partial \kappa}\left[\left(\frac{\partial U}{\partial \phi}\right)_T \left(\frac{\partial (\pi\sigma^3/6V]}{\partial V}\right)_T\right]\right)_V \left(\frac{\partial \kappa}{\partial T}\right)_V \quad (A6)$$

And, therefore,

$$\left(\frac{\partial C_V}{\partial \phi}\right)_T = \left(\frac{\partial}{\partial \kappa}\left[\left(\frac{\partial U}{\partial \phi}\right)_T\right]\right)_V \left(\frac{\partial \kappa}{\partial T}\right)_V \quad (A7)$$

Since $\left(\frac{\partial \kappa}{\partial T}\right)_V < 0$ for the Yukawa potential model of Eqn. (1). Also,

$$\left(\frac{\partial}{\partial \kappa}\left[\left(\frac{\partial U}{\partial \phi}\right)_T\right]\right)_V \sim -\left(\frac{\partial}{\partial \kappa}\left[\left(\frac{\partial U}{\partial r}\right)_T\right]\right)_V \quad (A8)$$

Then from differentiating Eqn. (1), one obtains

$$\begin{aligned}\left(\frac{\partial}{\partial \kappa}\left[\left(\frac{\partial U}{\partial r}\right)_T\right]\right)_V &= \left(\frac{\partial}{\partial \kappa}\left[\left(\frac{\partial}{\partial r}\{\Gamma\frac{\exp(-\kappa r)}{r}\right)_T\right]\right)_V \\ &= -\frac{\Gamma}{r}\left(\frac{\partial}{\partial \kappa}\left[(\frac{1}{r}+\kappa)\exp(-\kappa r)\right]\right)_V \\ &= -\frac{\Gamma}{r}\left(\exp(-\kappa r) - (1+\kappa r)\exp(-\kappa r)\right) = \Gamma\kappa r\exp(-\kappa r) > 0\end{aligned} \quad (A9)$$

Therefore,

$$\left(\frac{\partial C_V}{\partial \phi}\right)_T = \left(\frac{\partial}{\partial \kappa}\left[\left(\frac{\partial U}{\partial \phi}\right)_T\right]\right)_V \left(\frac{\partial \kappa}{\partial T}\right)_V > 0 \quad (A10)$$

Eqn. (A10) accounts for the basic trend of $C_V$ vs. $\phi$. Besides the analysis based on the pair interaction model of Eqn. (1), the behavior of $C_V$ is dependent on the number density of particles. Since more charged particles are needed to increase $\phi$, it increases degrees of freedom through an increase of pair interactions per unit volume. After all three models are transformed into crystal states, their $C_V$ decreases in the order of $r_s/\sigma = 0$, 0.3 and 0.1. These results can be understood as follows. When the charge site is shifted away from the center, the average potential energy increases (over all angles, as shown in the inset of Fig. 2), which leads to the greater effective size for the asymmetric particle of $r_s/\sigma$ = 0.1 compared to that of centrosymmetric particles. Consequently, asymmetric particles pack more tightly and their $Cv$ decreases due to suppression of spatial fluctuation. When $r_s/\sigma$ is increased from 0.1 to 0.3 at a given $\phi$ in the crystal state, the orientation fluctuation may contribute to a wider range of pair distances between the two charged sites for $r_s/\sigma$ = 0.3 for a given separation between two particles, which may also increase the range of possible potential energy values along with an increase of $Cv$.

**Appendix B. Calculation of Averaged Pair Potential Energy $<U(r)>$**

In our MC simulation, the average of pair potential energy is calculated based on the following procedures schematically displayed in Figure B1.

1. The distance between two particles is first fixed at $r$.

2. Then the charge is shifted by a distance $r_s$ from the center-of-mass toward the periphery.

3. The two particles are free to rotate. The equilibrium distance between two charge sites $r'$ is registered and the pair potential energy is calculated.

4. The calculation is repeated one millions times. The potential energy of each step is summed up and divided by the total number of steps to obtain $<U(r)>$.

**For Table of Contents Use Only**

Luis E. Sánchez-Díaz, Chwen-Yang Shew*, Xin Li, Bin Wu, Gregory S. Smith and Wei-Ren Chen*

Phase Behavior under a Non-Centrosymmetric Interaction: Shifted Charge Colloids Investigated by Monte Carlo Simulation

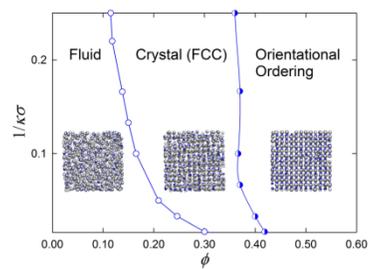

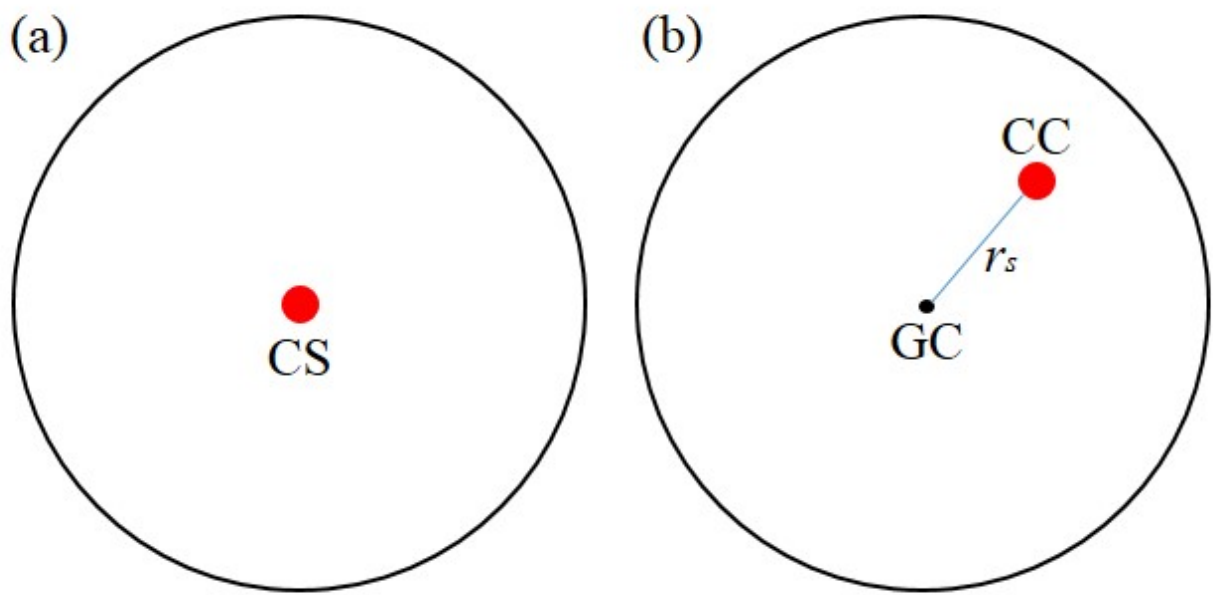

**Figure 1.** Schematic representation of position of charge. The left side is the case when the charge is in the center. The right side is when the charge is shifted by a distance $r_s$ from of the center. The label CS (centro symmetry), GC (geometric center) and CC (center of charge) is the coordinates we use to specify the relevant positions.

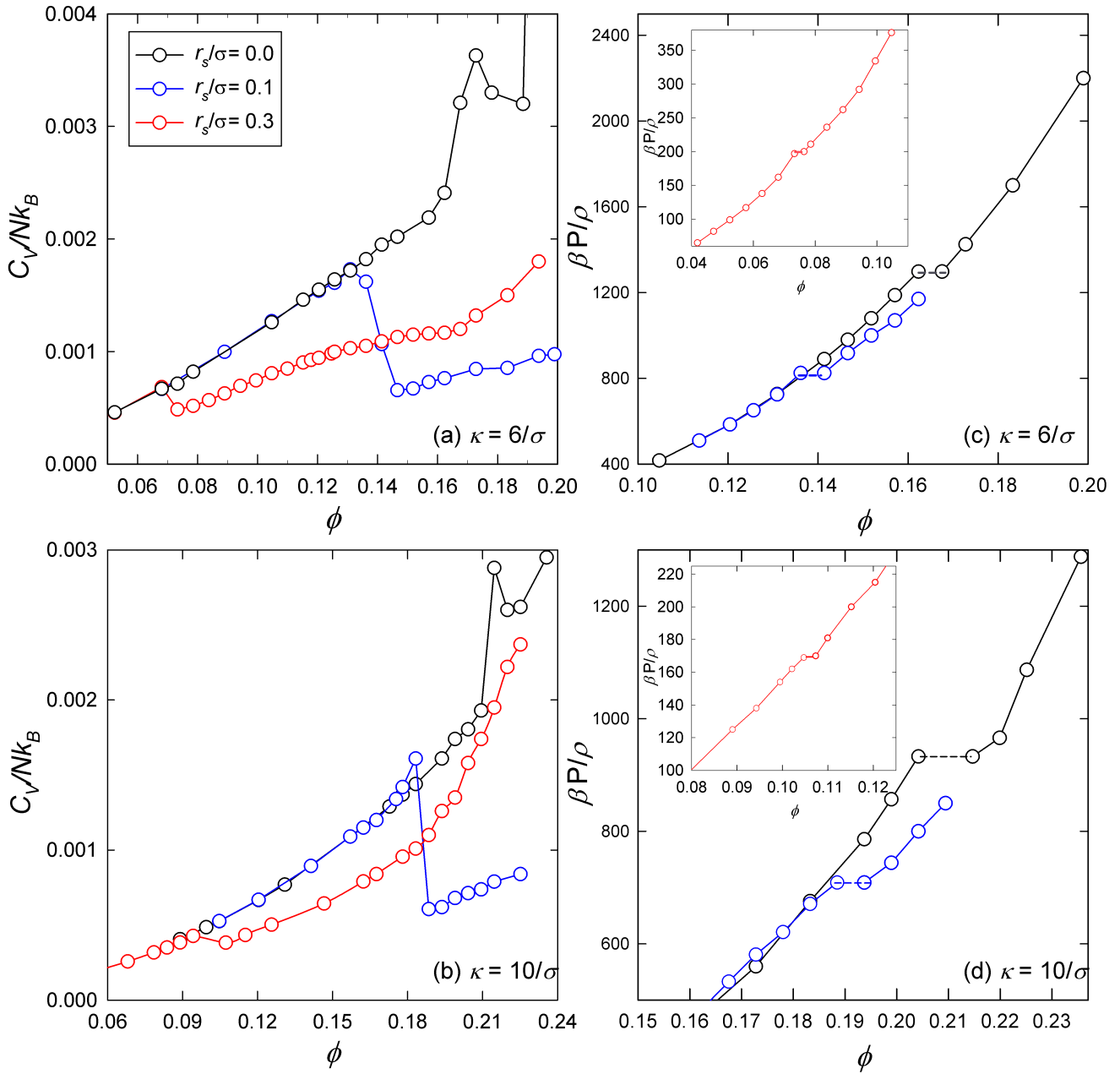

**Figure 2.** Heat capacity ((a): & (b)) at constant volume $Cv$, and pressure ((c) & (d)) as a function of colloidal volume fraction for particle with different values of $r_s$. The lines connecting the data points are a guide for the eyes. The value of $\kappa$ is $6/\sigma$ for (a) and (c) and is $10/\sigma$ for (b) and (d). In both calculations, the value of $\Gamma$ is 81 $k_B T$.

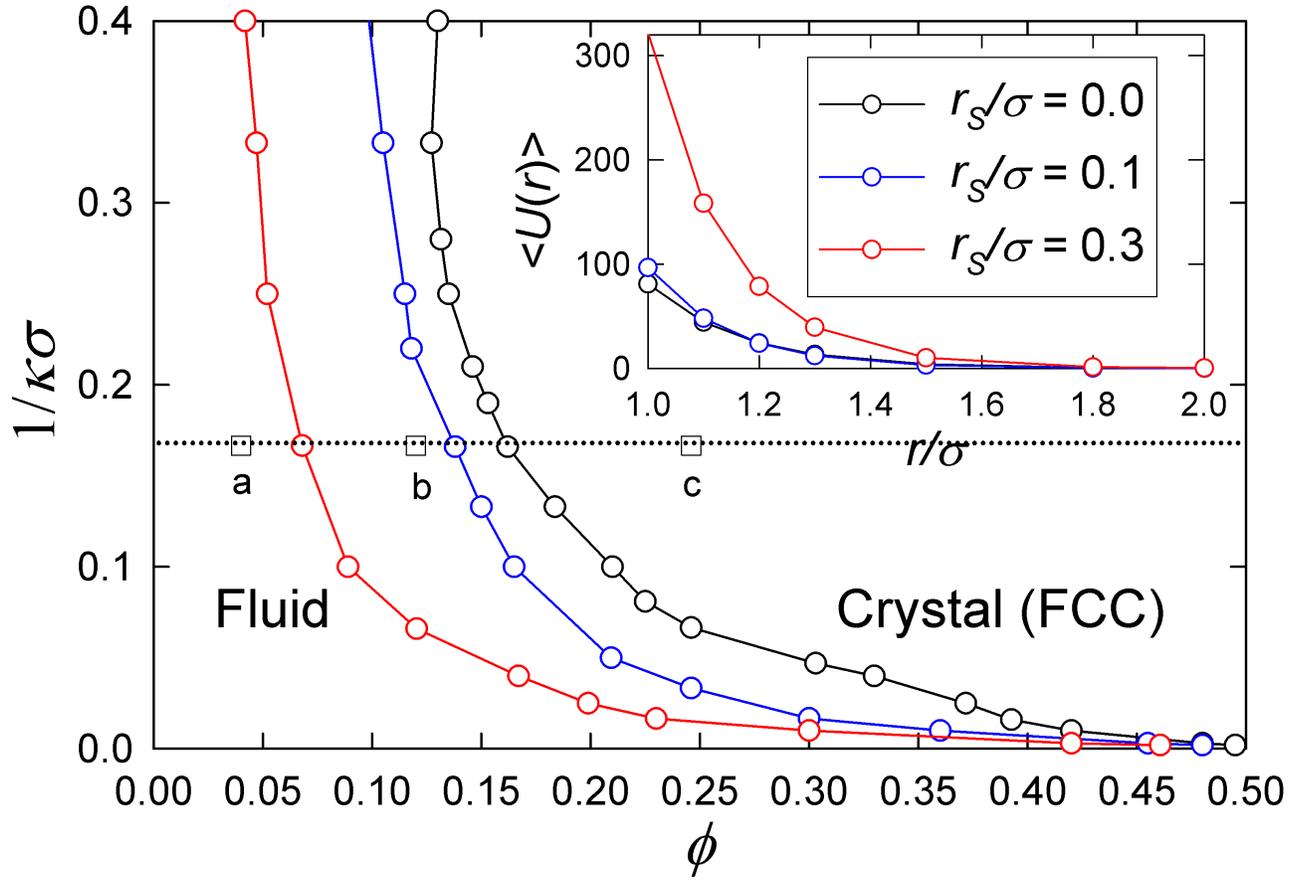

**Figure 3.** The MC-calculated phase diagram of particles with shifted charge in the $1/\kappa\sigma$ - $\phi$ plane. The magnitude of Yukawa repulsion $\Gamma$ is 81 $k_\text{B}T$. The dotted line gives the constant value of 1/6 for $1/\kappa\sigma$. In the inset, we show the average pair potential energy $<U(r)>$ with different values of $r_s$.

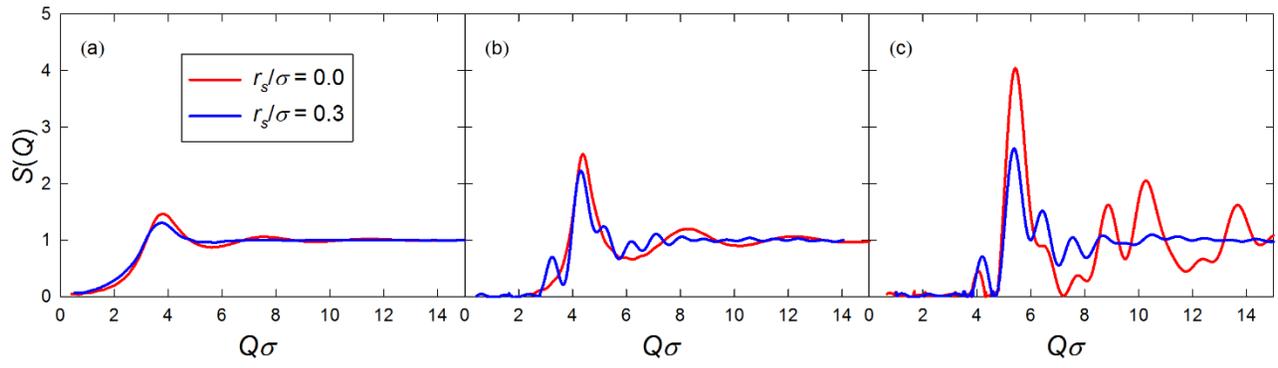

**Figure 4.** The MC-calculated inter-particle structure factor $S(Q)$ for states a, b and c along the dotted line gives in Figure 3.

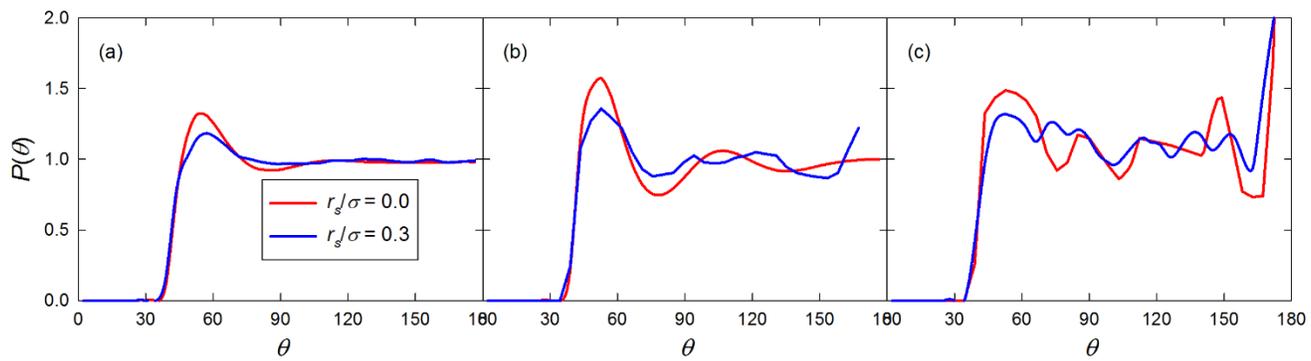

**Figure 5.** The MC-calculated bond angle distribution function (BADF) $P(\theta)$ for states a, b and c along the dotted line gives in Figure 3.

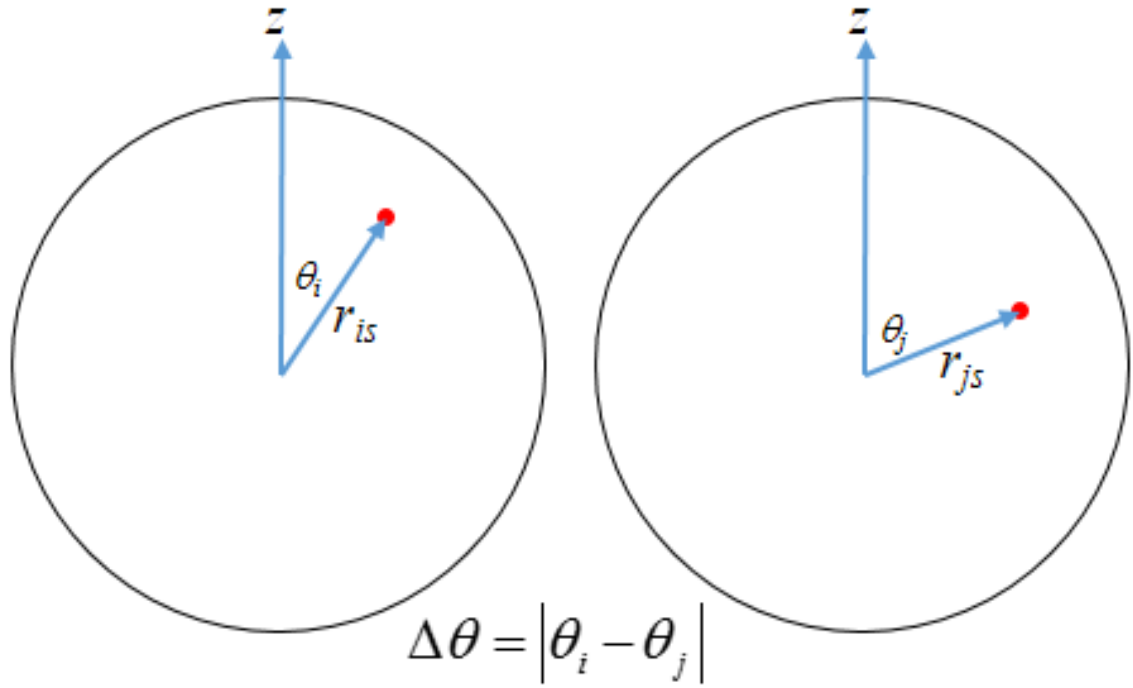

**Figure 6.** Definition of the polar angle $\Delta\theta$ used in Eqn. (4).

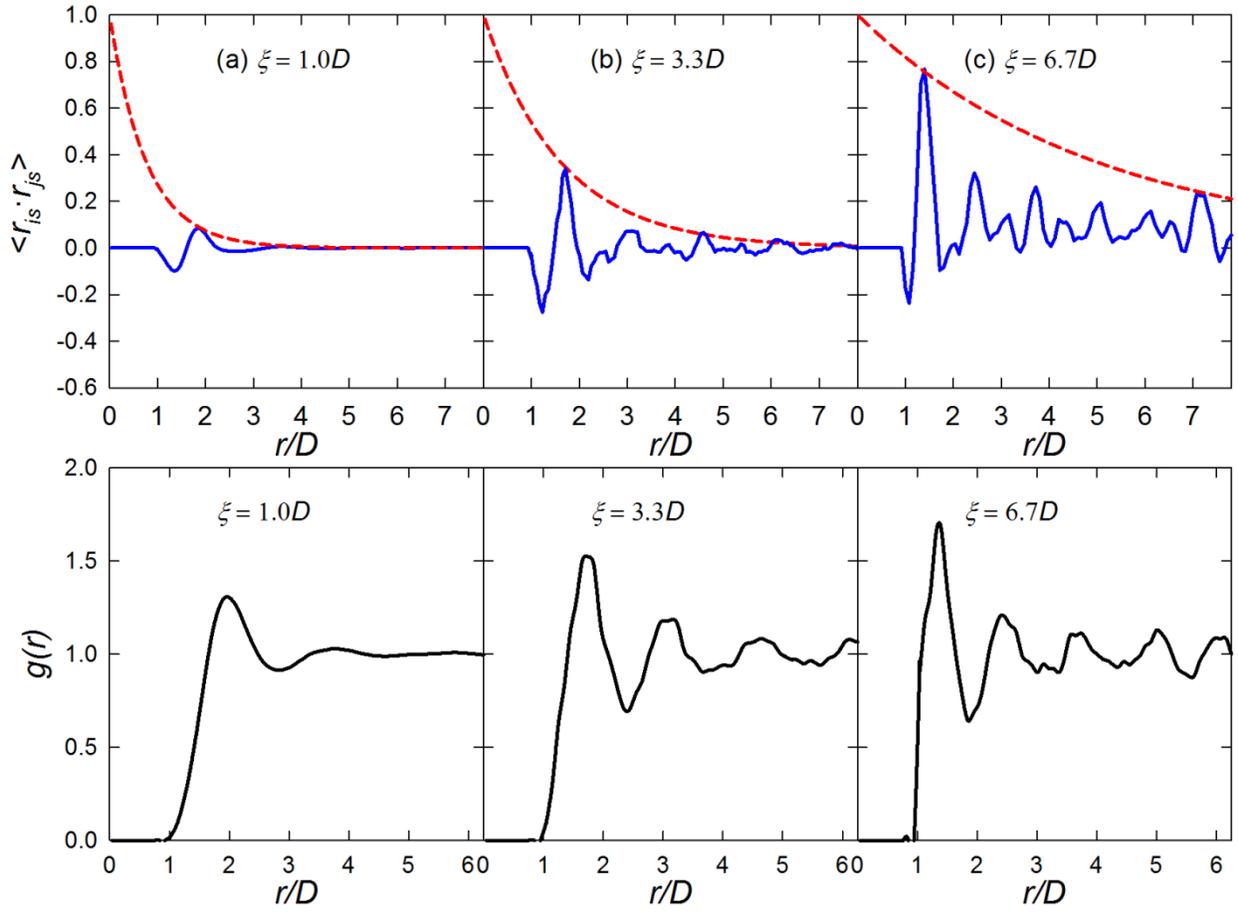

**Figure 7.** Procedure to calculate correlation length $\xi$ (top) and the corresponding radial distribution function $g(r)$ (bottom) for the three states a, b, c of particles with $r_s = 0.3$ shown in Figure 3. $D$ is the particle diameter. In the top, the procedure to calculate the correlation length $\xi$ using Eqn. (6) (blue solid lines). It is clearly seen that the oscillations of $\langle \vec{r}_{is} \cdot \vec{r}_{js} \rangle$ is due to the weighting of $g(r)$ in the calculation. We used Eqn. (5) to fit the peaks of $\langle \vec{r}_{is} \cdot \vec{r}_{js} \rangle$.

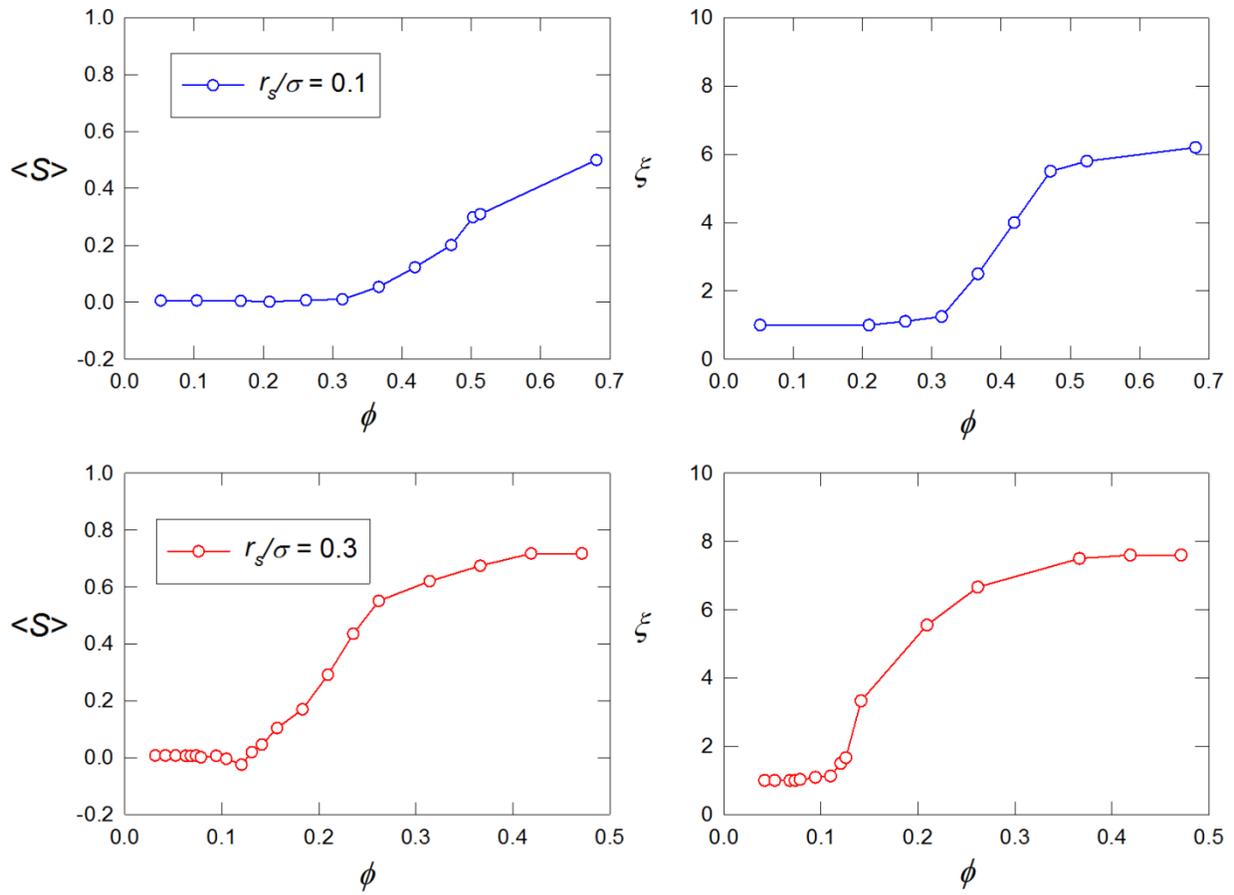

**Figure 8.** The order parameter <S> and correlation length ξ as a function of colloidal volume fraction $\phi$ for particle with different values of $r_s$.

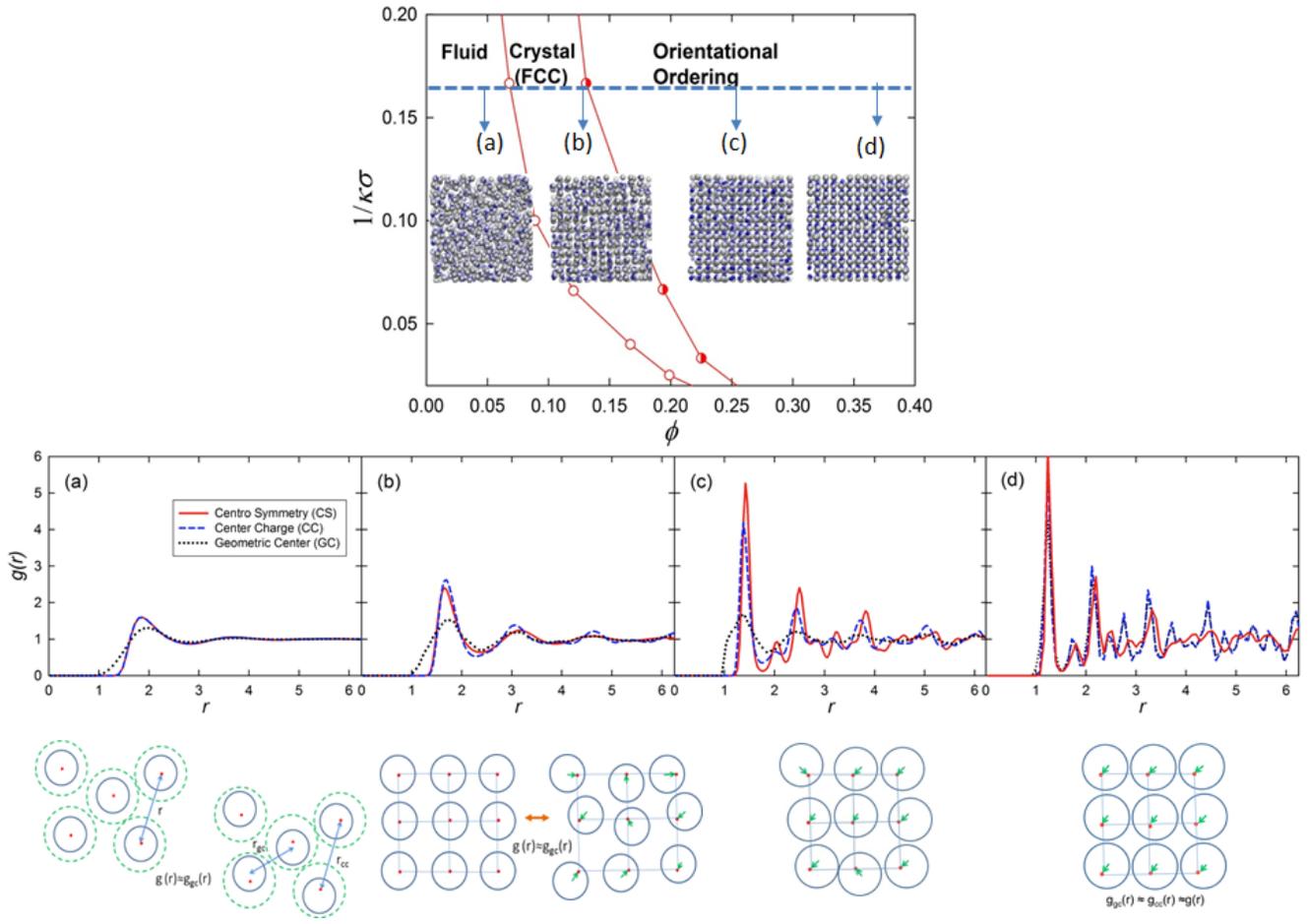

**Figure 9.** Plot of the phase diagram of asymmetric particles ($r_s/\sigma = 0.3$) with typical snapshots when $\Gamma = 81$ and $1/\kappa\sigma = 1/6$ on top panel, where charge sites are labeled with blue dots, and comparison of the pair correlation function g(r) between centrosymmetric particles ($r_s/\sigma = 0$) with the $g_{cc}$(r) (between charge sites) and $g_{gc}$(r) (between geometric centers) of asymmetric particles for $\phi = 0.05$(a), 0.13(b), 0.25(c), and 0.366 (d) on middle panel along with two-dimensional schematics to account for the corresponding structures on bottom panel.

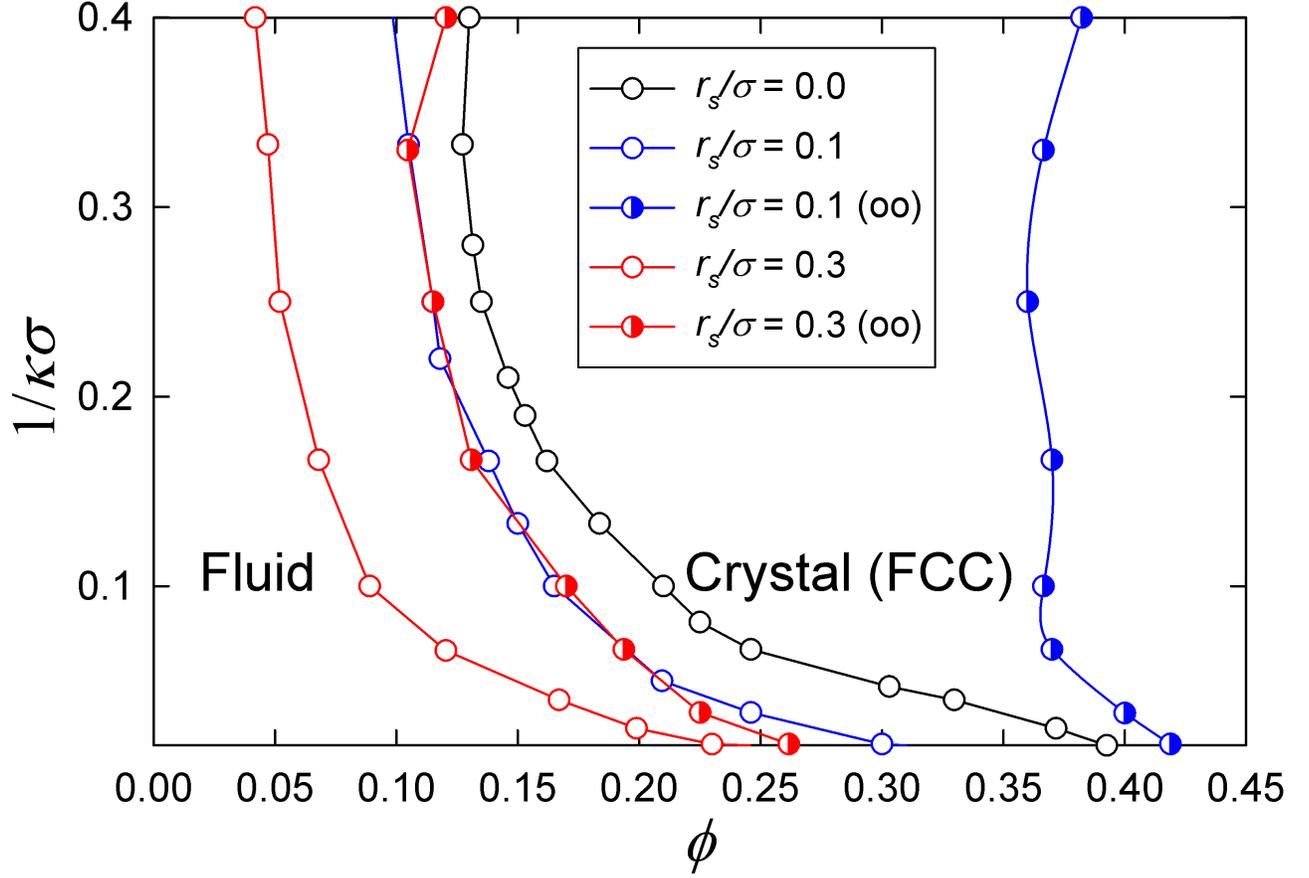

**Figure 10.** The MC-calculated phase diagram with orientational ordering transition of particles with shifted charge in the $1/\kappa\sigma$ - $\phi$ plane. The magnitude of Yukawa repulsion $\Gamma$ is 81 $k_B$T. In contrast to centrosymmetric particles, asymmetric particles undergo two transitions as the packing fraction is increase. First, we observer the transformation from liquid to crystal (open circles), and a further increase of packing fraction induces the transition of orientation ordering (semi-open circles). As the particle becomes more asymmetric ($r_s/\sigma$ increased from 0.1 to 0.3), the phase boundaries shifts towards smaller packing fraction.

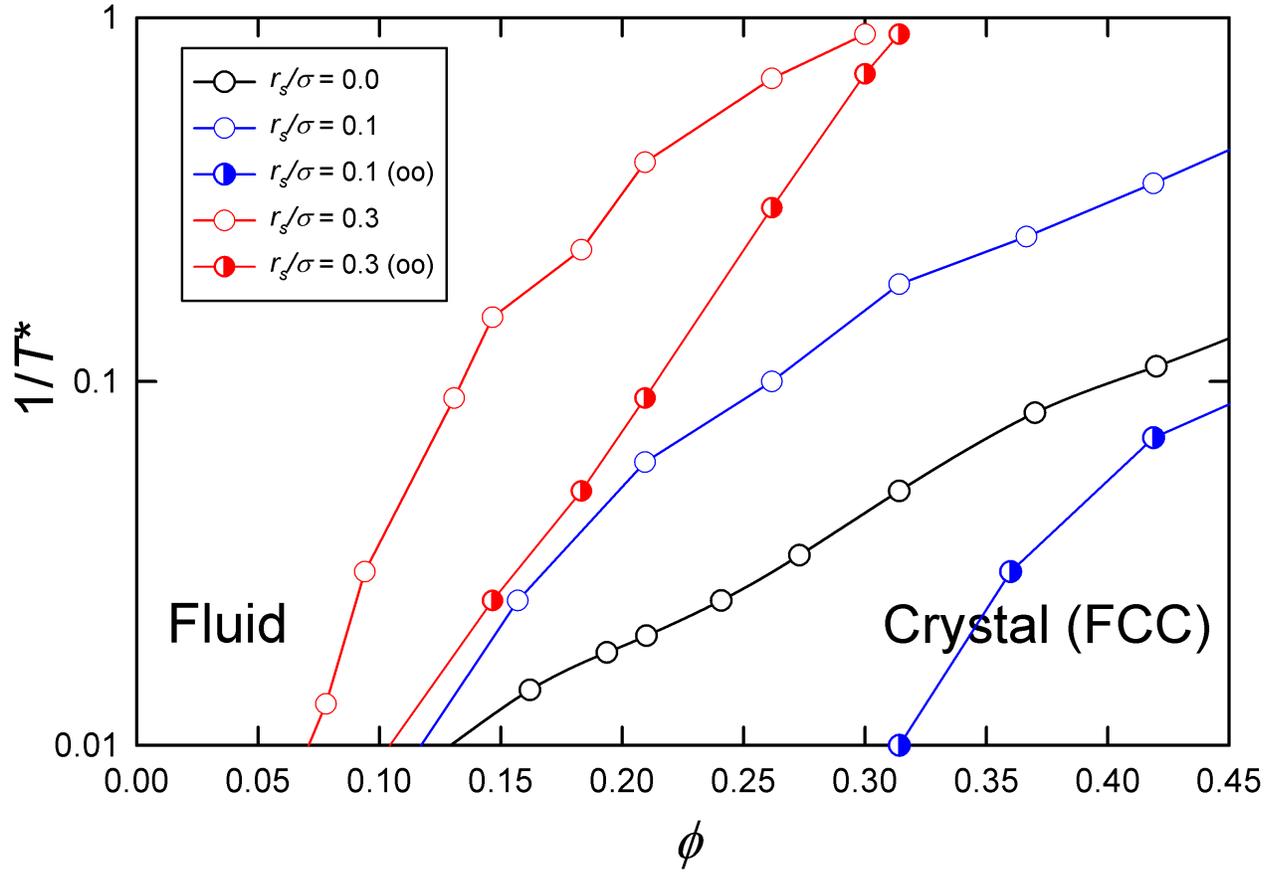

**Figure 11.** The MC-calculated phase diagram of particles with shifted charge in the $1/T^*$-$\phi$ plane. The magnitude of screening parameter $\kappa\sigma$ is 6. The argument of the competition between the potential energy and configurational entropy can also be used to explain the temperature dependence of the phase boundary.

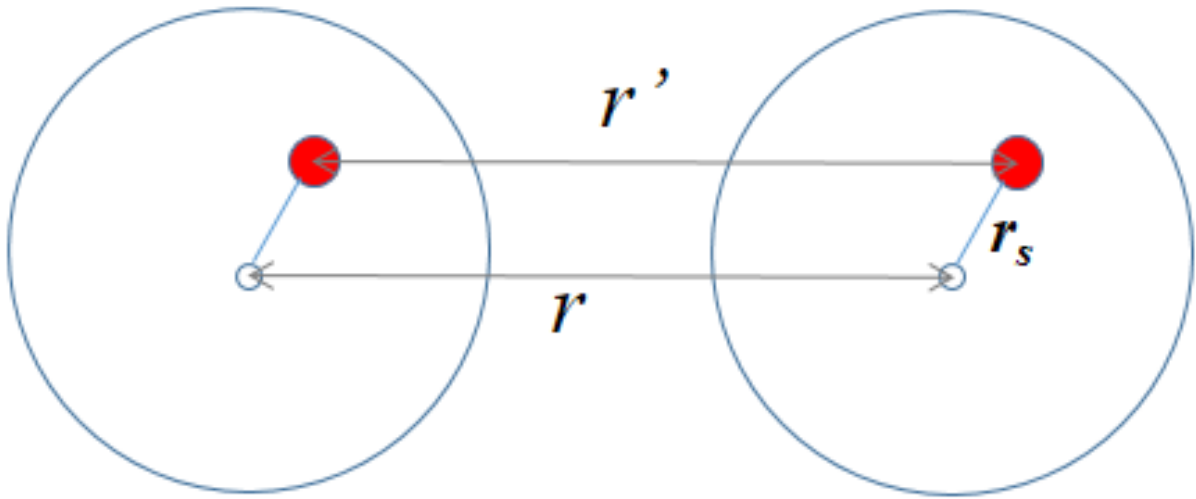

**Figure B1.** Schematic representation of the interaction $u(r)$ between two particles when charge is shifted a distance $r_s$. $r$ is the distance between the geometrical center and $r'$ is the distance between the positions of two different charges.